\documentclass[twocolumn]{aa} 

\usepackage{graphicx}
\usepackage{txfonts}
\usepackage{amsmath}

\begin{document}

   \title{An unusual type-I X-ray burst from the neutron star X-ray binary IGR J17591--2342: a double-photospheric-radius-expansion burst?}

\author{Sudip Bhattacharyya
\inst{1}
\and
Akshay Singh
\inst{2}
\and
Andrea Sanna
\inst{3,4}
}

\institute{Department of Astronomy and Astrophysics, Tata Institute of Fundamental Research, 1 Homi Bhabha Road, Colaba, Mumbai 400005, India;
\email{sudip@tifr.res.in}
\and
Department of Physics, Bar-Ilan University, Ramat-Gan 52900, Israel;
\email{akshaysingh.astro@gmail.com}
\and
Dipartimento di Fisica, Universit\'a degli Studi di Cagliari, SP Monserrato-Sestu, KM 0.7, Monserrato,I-09042, Italy
\and
INFN, Sezione di Cagliari, Cittadella Universitaria, I-09042 Monserrato, CA, Italy; 
\email{andrea.sanna@dsf.unica.it}
}

   \date{Received xxxx; accepted xxxx}

\titlerunning{An unusual type-I X-ray burst from IGR J17591--2342}

\abstract
{Type-I X-ray bursts observed from neutron stars 
originate from intermittent unstable
thermonuclear burning of accreted matter on these stars.
Such bursts, particularly those reaching the Eddington luminosity
and having a temporary photospheric radius-expansion due to
radiation pressure, provide a testbed to study nuclear fusion processes
in intense radiation, gravity, and magnetic fields.}
{Here, we investigate time-resolved spectroscopic properties of a type-I burst
from the accretion-powered millisecond X-ray pulsar IGR J17591--2342.}
{Our basic spectral model includes an absorbed blackbody to describe the burst 
emission and an absorbed power law to represent the non-burst emission.}
{The blackbody normalisation shows two consecutive humps aligned with
blackbody temperature dips during the burst. Such an unusual behaviour 
could imply two consecutive photospheric radius-expansion
events during the same burst or
a systematic metallicity evolution in the neutron star atmosphere.  
However, our spectral analysis suggests the latter option is less likely to be happening for IGR J17591--2342.}
{The novel former option implies that sufficient fuel survived after the first photospheric radius-expansion event to power a second similar event a few seconds later, challenging the current theoretical understanding.
If confirmed, the double photospheric radius-expansion event observed in IGR J17591--2342 suggests the possibility of avoiding photospheric expansion at luminosities exceeding Eddington.
Mechanisms such as temporary enhancement of the magnetic field by convection and confinement of the plasma could be invoked to explain the peculiar behaviour of the source.}


\maketitle
%

\section{Introduction} \label{introduction}

A neutron star low-mass X-ray binary (LMXB) primarily emits X-rays due to the mass transfer from a low-mass companion
star to a neutron star via an accretion disc \citep{Bhattacharyaetal1991}.
This accreted matter or fuel accumulates on the stellar surface,
and can intermittently burn via thermonuclear runaway fusion processes,
giving rise to a burst of radiation observed in X-rays, typically for several tens
of seconds \citep[][and references therein]{Grindlay1976,LambLamb1978,StrohmayerBildsten2006,Gallowayetal2020}.
Such a burst, known as a type-I X-ray burst, is expected to ignite at a certain point on the star, and then the thermonuclear flames may spread to engulf the entire surface, typically, during the early part of the burst rise
\citep{Spitkovskyetal2002,BhattacharyyaStrohmayer2007,ChakrabortyBhattacharyya2014}.
X-ray intensity variations with frequencies close to the neutron star's spin frequency are detected for some bursts. These burst oscillations may originate
due to an azimuthally asymmetric brightness pattern on the spinning 
stellar surface \citep{Watts2012,Bhattacharyya2022}.
The spectral and timing aspects of thermonuclear bursts can be useful to probe the superdense matter in the neutron star core 
\citep[][and references therein]{Bhattacharyya2010} and plasma and nuclear 
physics in the strong gravity regime \citep{StrohmayerBildsten2006}.

It is particularly important to study some of these topics for the most
intense bursts, e.g., the photospheric radius-expansion (PRE) bursts,
for which the burst luminosity temporarily exceeds the
Eddington luminosity
\citep{ShapiroTeukolsky2004,Gallowayetal2020,Strohmayeretal2019}.
For this luminosity, the outward radiative force temporarily exceeds the inward gravitational force,
pushing the stellar photosphere away from the star.
Note that the identification and study of these phenomena require the tracking of the evolution 
of burst spectral and timing properties using a time-resolved analysis
\citep[e.g., ][]{BhattacharyyaStrohmayer2006,Bultetal2019}.
Such an analysis finds that the temperature of the photosphere decreases 
as it expands, and increases again as the photosphere contracts
and touches the neutron star surface. 

This paper reports the results of time-resolved spectroscopy applied to a peculiar type-I X-ray burst observed from the neutron star LMXB IGR J17591--2342. 
This source is an accretion-powered millisecond X-ray pulsar 
(AMXP) with the spin frequency of $\approx 527.4$ Hz,
for which the neutron star magnetosphere channels the accreted matter to stellar magnetic poles \citep{Sannaetal2018, Bhattacharyaetal1991}.
\citet{Kuiperetal2020} reported a type-I burst from this source.
They concluded that this was most likely
a PRE burst from its fast rise ($\sim 1-2$~s) and a subsequent 
$\sim 10$ s long plateau at its peak, but without involving 
time-resolved spectroscopy.
Here, the study using time-resolved spectroscopy of another type-I burst 
observed earlier with {\it AstroSat} suggests an unusual nature, viz., two likely consecutive PRE events within a few seconds during the same burst.
An interpretation of this novel scenario concerns two of the most crucial 
questions regarding such type-I X-ray bursts---if the available magnetic field can confine the plasma during
a nuclear fusion, and if a significant amount of fuel in the vicinity
of the burning region can remain unburnt.

We mention the observation of IGR J17591--2342 in section~\ref{observation}, 
describe colour and preburst spectral analysis in section~\ref{color},
detail burst spectral analysis in section~\ref{burst}, and discuss the results and interpretation in section~\ref{Discussion}.

\section{Observation}\label{observation}

\begin{figure}[ht]
\centering
\hspace{-1.0cm}
\includegraphics*[width=9cm,angle=0]{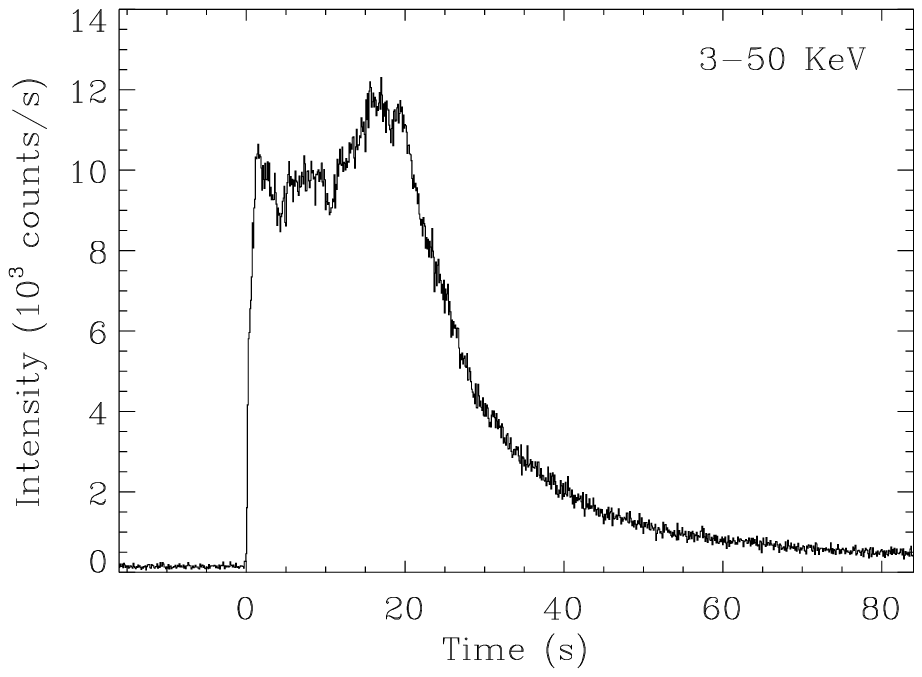}
        \caption{ 
The X-ray ($3-50$ keV) intensity
versus time (with 0.1 s bins) of a thermonuclear type-I burst observed from the accreting millisecond X-ray pulsar IGR J17591--2342 with
{\it AstroSat}/LAXPC (section~\ref{observation}).
\label{fig1}}
\end{figure}

The AMXP IGR J17591--2342 was observed
with {\it AstroSat} Large Area X-ray Proportional
Counters\footnote{https://www.tifr.res.in/$\sim$astrosat\_laxpc/astrosat\_laxpc.html}
\citep[LAXPC; ][]{Antiaetal2017, Antiaetal2021} in Event Mode
from 2018-08-23T01:03:50 to 2018-08-24T00:35:00 UTC
(Observation ID (ObsID):
9000002320).
The data are, however, not continuous, because {\it AstroSat} is
a low Earth orbit satellite, and there are Earth occultation and
South Atlantic Anomaly passages
\citep{SeethaKasturirangan2021}.
The energy range, area and time resolution of LAXPC are
$\approx 3-80$ keV, $\sim 6000$~cm$^2$ and $\approx 10$ $\mu$s, respectively
\citep{Antiaetal2017}.
We use one LAXPC detector, viz., LAXPC20,
because the other two detectors had leak and gain stability issues
\citep{Antiaetal2021}. Using the laxpc3.4.3 software, we analyse the LAXPC data, extract light curves, spectra,
background and response files,
and correct the effects of deadtime in light curves and spectra.
Note that blank sky observations are used to estimate the LAXPC background.
We identify a type-I X-ray burst due to its typical sharp rise,
top plateau-type portion and near-exponential decay (see Figure~\ref{fig1}).
This burst, which occurred on August 23, 2018, started around 03:38:31 UTC.
In this paper, we probe and interpret the spectral evolution of this burst.

Here, we note that a recent paper \citep{Mancaetal2023}
mentioned an unexpected spectral nature
of the source during the ObsID 9000002320, which, they guessed, could have
been due to ``a bad calibration''. However, we find that such a spectral nature was due to a large intensity increase observed with
LAXPC during a part of this ObsID \citep[see also][]{Singhetal2025}.
This increase was not due to the X-ray intensity increase
of IGR J17591--2342, as we confirm from a few short observations
with the {\it NICER} satellite \citep{Mancaetal2023}
conducted during the {\it AstroSat} ObsID 9000002320.
Note that there was no {\it NICER} observation during the type-I X-ray burst.
The LAXPC count rate increase was due to a drift of
the satellite pointing from IGR J17591--2342 towards a nearby bright
persistent source GX 5--1 \citep[confirmed from a discussion with the
LAXPC team; see also][]{Singhetal2025}.
But the type-I X-ray burst occurred during the initial part of the ObsID 9000002320, when the LAXPC count rate increased had
not begun yet, and the observed spectrum was consistent with the expected source spectrum. Therefore, our burst
and preburst spectra were genuinely from IGR J17591--2342 and without
any calibration problem.

\section{Colour estimation and preburst spectral analysis}\label{color}

\begin{figure}[ht]
\centering
\includegraphics*[width=8cm,angle=0]{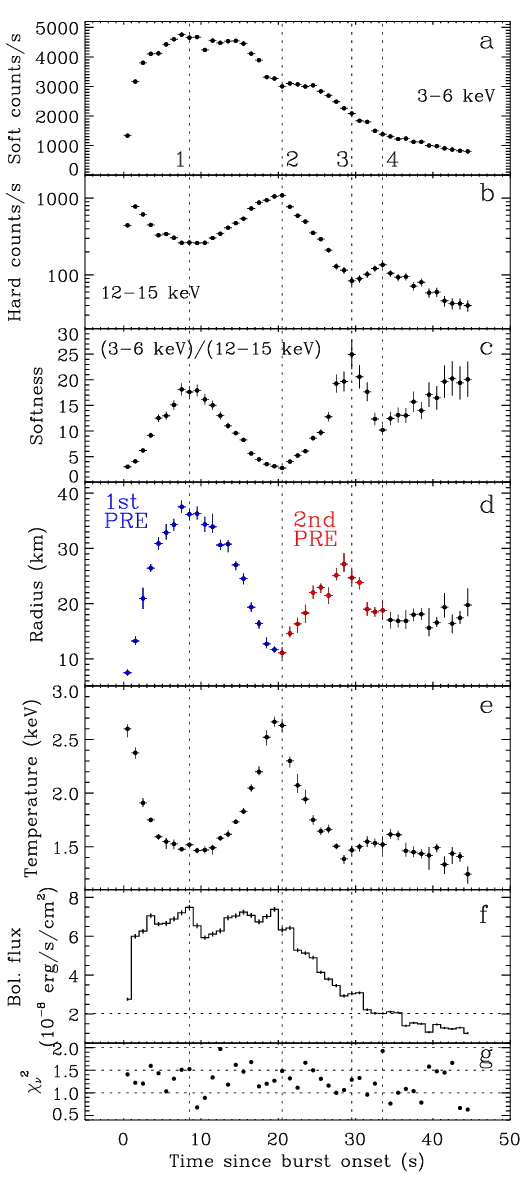}
        \caption{Time-resolved (1 s bins) properties with 1$\sigma$
        errors of the thermonuclear
        X-ray burst observed from IGR J17591--2342 with {\it AstroSat}/LAXPC.
         {\it Panels a--c}: Background-subtracted count rates in the energy range $3-6$ keV, $12-15$ keV, and their ratio, respectively.
        {\it Panels d--f}: Blackbody radius ($r_{\rm BB}/D_{10}$), blackbody temperature
        $kT_{\rm BB}$, and bolometric (0.01--100 keV) blackbody flux $F_{\rm BB}$ from the burst intervals best-fits
       using the model {\tt tbabs(bbodyrad+constant*powerlaw+gauss+gauss)} (see sections~\ref{color}, \ref{burst} and \ref{Discussion} for more details).
        The dotted horizontal line marks the $F_{\rm BB}$ at
        the second touchdown point (see section~\ref{flux}).
        {\it Panel g}: Reduced $\chi^2$ of spectral fits.}
\label{fig2}
\end{figure}

In order to study the evolution of burst properties, we divide the first 45 seconds of the burst into 1-second time bins.
First, we study the colour evolution during the burst, which gives a model-independent overview of the source spectral evolution, providing a sanity check for spectral results.
We define the colour, or softness, as the background-subtracted count rate ratio between $3-6$ keV (channels $5-10$) and $12-15$ keV (channels $22-27$).
Panels~{\it a--c} of Figure~\ref{fig2} show the profiles of $3-6$ keV
count rate, $12-15$ keV count rate, and softness.
The softness has two clear consecutive humps followed by an increase, which
suggest a preliminary softening of the source spectrum followed by two hardening phases, after which the source continued to soften during the rest of the burst.
 
In line with the standard relatively short pre-burst interval selection \citep{Jaisawaletal2019}, we generate a pre-burst spectrum for the 200 s interval just before the burst onset, ensuring at least 25 counts in each bin. We model the LAXPC non-burst spectrum of IGR J17591--2342 with an absorbed Comptonization model ({\tt tbabs*Nthcomp} in  {\it XSPEC}), in line with the detailed analysis performed on a broader energy range combining multiple available datasets \citep[see ][for more details]{Mancaetal2023}. Due to the lack of coverage below 3 keV, we do not include an edge kept fixed at 0.871 keV by \cite{Mancaetal2023}, and freeze the hydrogen column density at $2.25\times10^{22}$ cm$^{-2}$ \citep[][]{Kuiperetal2020}. Finally, we include a 3\% systematic uncertainty for LAXPC spectral fitting \citep{Maqbooletal2019}.

The reduced $\chi^2$ ($\chi^2_{\nu} = \chi^2$/DoF; DoF is degrees of freedom) is $\approx 1.55$ ($51/33$) and the absorbed flux is
$\sim 3\times10^{-10}$~erg~s$^{-1}$~cm$^{-2}$ for the {\it XSPEC} model {\tt tbabs*Nthcomp} fit in the $3-25$ keV range.
The addition of a blackbody component does not improve
the fit. This aligns with \cite{Mancaetal2023}, which only sporadically detected a significant blackbody contribution during the outburst, characterised by a temperature of $0.8\pm0.2$~keV, outside the LAXPC energy range.
We note that the addition of a Gaussian around 6.6 keV marginally improves the $\chi^2_{\nu}$(DoF) to $\approx 1.38(30)$. However, since a Gaussian component in the $\sim 6-8$ keV range will be included to describe the burst spectrum, we decide not to include it for the pre-burst modelling. 

We also explore if an absorbed power law model ({\tt tbabs*powerlaw} in {\it XSPEC}) could adequately describe the pre-burst spectrum. We obtain a $\chi^2_{\nu}$(DoF) of $\approx 1.44(35)$, for the best-fit power law photon index and
normalization of $2.10\pm0.04$ and $0.11\pm0.01$, respectively. Since the power law component mimics the Comptonization well, we use it to describe the pre-burst emission for most spectral fits. In addition to ensuring a similar $\chi^2$ value
for a lesser number of free parameters (compared to that for {\tt tbabs*Nthcomp}), the {\tt tbabs*powerlaw} model allows us to explore the effects of an evolving non-burst spectral shape only by keeping free a single parameter (the power law photon index), while for {\tt Nthcomp} several free parameters are required. Nevertheless, for completeness, we also test the burst spectral fits using {\tt Nthcomp} to describe the non-burst spectrum.

\section{Burst spectra: analysis, results and interpretation}\label{burst}

After characterising a pre-burst spectrum of IGR J17591--2342,
we model the spectra of the first 45 independent burst time bins, each of 1 s 
(and also the first 22 independent burst time bins, each of 2 s, to show that the conclusions do not depend on the choice of the bin size).
Similar to the pre-burst spectral analysis (see section~\ref{color}), we ensure that there are
at least 25 counts in each spectral bin, use 3\% systematics, freeze the hydrogen column density at $2.25\times10^{22}$ cm$^{-2}$
and fit in $3-25$ keV range ($3-12$ keV only for the third 1 s time bin, which shows systematics at higher energies).

\subsection{Basic model}\label{Basic_model}

The burst spectra are usually fit with a blackbody ({\tt bbodyrad} in XSPEC) model \citep{Gallowayetal2008}, which has two parameters---temperature ($kT_{\rm BB}$; $k$ is the Boltzmann constant) and normalization ($N_{\rm BB} = r_{\rm BB}^2/D_{10}^2$).
Throughout this paper, we call $r_{\rm BB}/D_{10}$ the blackbody radius, where $D_{10}$ is the source distance in units of 10 kpc.
The blackbody effective temperature of the neutron star photosphere ($kT_{\rm eff}$) shifts to a higher value (colour temperature $kT_{\rm c}$) due to electron scattering in the stellar atmosphere \citep{Londonetal1986}, and hence, the blackbody here is diluted.
The effective temperature $kT_{\rm eff}$ is inferred by rescaling the observed temperature $kT_{\rm BB}$ by the factor $(1+z_r)/f_{\rm c}$, where $f_{\rm c}$ is the colour factor and $1+z_r = [1 - (2GM/rc^2)]^{-1/2}$ for Schwarzschild spacetime, where $G$ and $M$ are the gravitational constant and neutron star mass, respectively \citep{ShapiroTeukolsky2004,Damenetal1990}. Similarly, the photospheric radius $r$ is inferred by rescaling $r_{\rm BB}$ by the factor $f_{\rm c}^2/(1+z_r)$.

\begin{figure}[h]
\centering
\hspace{-1.0cm}
\includegraphics*[width=8cm,angle=0]{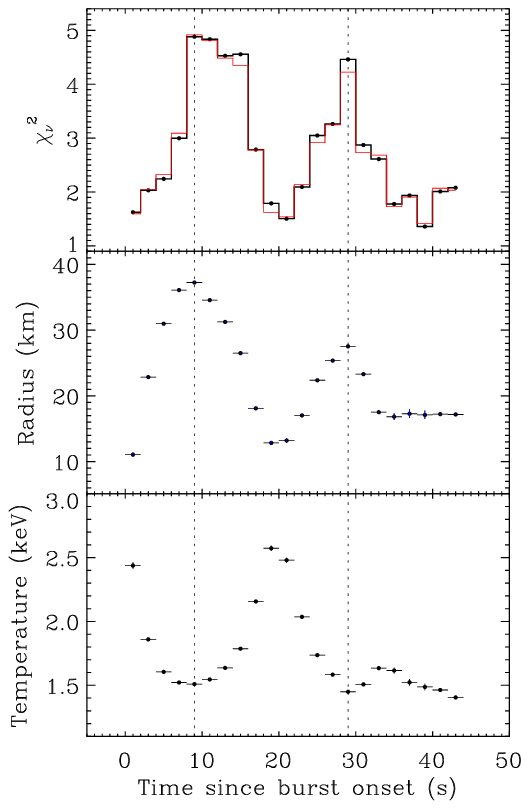}
        \caption{Time-resolved (2 s bins) properties, with $1\sigma$
errors, of the thermonuclear X-ray burst observed from IGR J17591--2342
with {\it AstroSat}/LAXPC 
(see section~\ref{Basic_model}).
The {\it upper panel} shows reduced $\chi^2$ profiles for spectral fitting with
the {\it XSPEC} models {\tt tbabs(bbodyrad+constant*powerlaw)} (black;
power law parameter values frozen to the pre-burst values), and
{\tt tbabs(bbodyrad+powerlaw)} (red). This panel shows that fitting is not acceptable; overall, fitting does not improve when the spectral shape (power law photon index) is made free, and
reduced $\chi^2$ profiles have two humps.
The {\it middle} and {\it lower panels} show blackbody ({\tt bbodyrad}) parameter
(radius and temperature, respectively) profiles for the first model mentioned above. These panels (along with the dotted
vertical lines) show that radius humps and temperature dips coincide
with reduced $\chi^2$ humps.
}
\label{figS0}
\end{figure}

In order to account for both the pre-burst and burst components, we use the {\it XSPEC} model
{\tt tbabs(bbodyrad+constant*powerlaw)}, which should be considered the basic model for burst spectra. The power law component has frozen pre-burst best-fit values parameters, while the free energy-independent constant allows for possible flux variations of the pre-burst emission during the burst. Note that, while this is the standard way to describe the non-burst emission during the burst \citep[see, e.g.,][]{Jaisawaletal2019}, we also perform fits
with a free power law photon index parameter to investigate a possible evolution of the pre-burst spectral shape.
Furthermore, while an absorbed power law adequately describes the pre-burst spectrum (as shown in section~\ref{color}), we also replace it with a Comptonization model component to check the reliability of our results.

For the fitting with the {\tt tbabs(bbodyrad+constant*powerlaw)} model,
we obtain high and unacceptable $\chi^2_{\nu}$ values for most of the
intervals (e.g., upper panel, Figure~\ref{figS0}).
This suggests that the basic model does not adequately describe the data.
Note that two humps of the burst blackbody radius ($r_{\rm BB}/D_{10}$) 
profile are clearly seen in Figure~\ref{figS0}, which are aligned with
the blackbody temperature ($kT_{\rm BB}$) dips. 
Interestingly, the two $\chi^2_{\nu}$ humps align with two humps of radius and two dips of temperature of the blackbody component.
For example, for the above-mentioned basic model, 
$\chi^2_{\nu}$(DoF) at the burst onset (1.62(34))
monotonically increases to 4.88(29) after about 9 s, decreases, and reaches a minimum of 1.51(34) after about 21 s since the burst onset, monotonically increases again to 4.46(22) after about 29 s since the burst onset, and then decreases. 
This shows that the basic spectral model is particularly inadequate during the two humps of radius and two dips of temperature of the blackbody component.
We test whether the inadequacy of the basic model derives from an evolution of the power law photon index; however, leaving the photon index free did not significantly improve the fits (Figure~\ref{figS0}).

\begin{figure}[ht]
\centering
\hspace{-1.0cm}
\includegraphics*[width=8.5cm, angle=0]{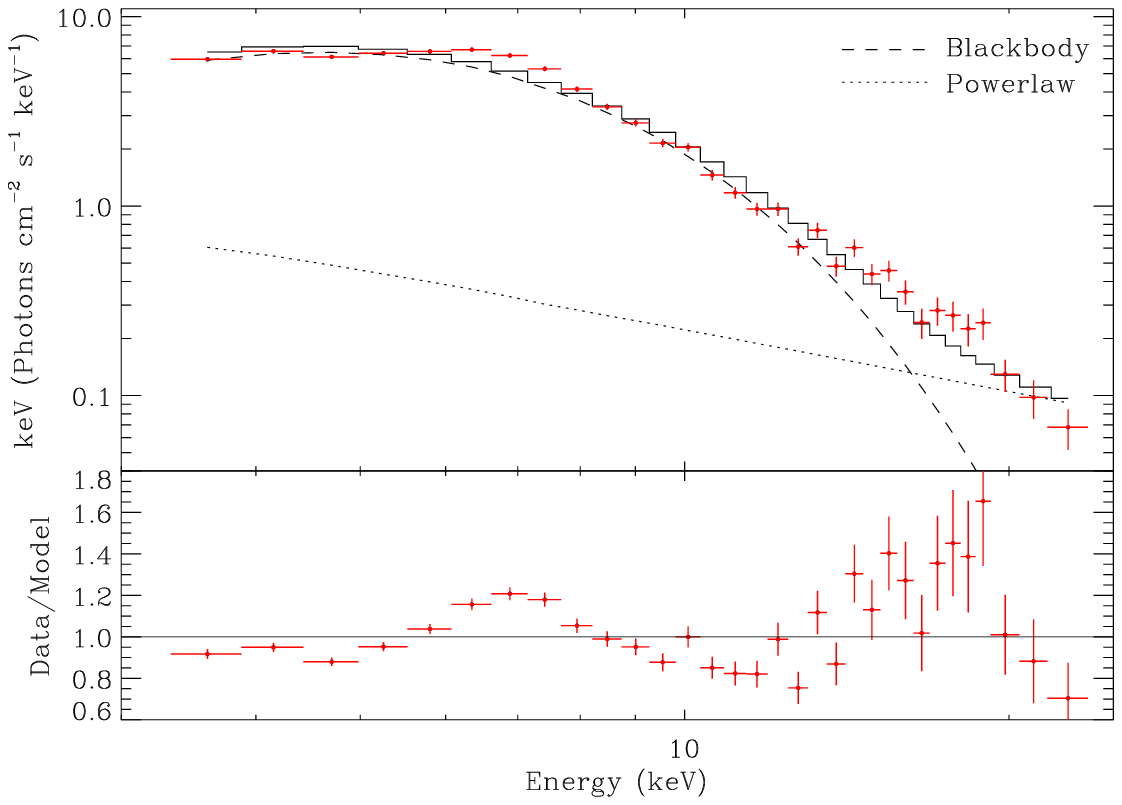}
        \caption{Spectral fitting for the thermonuclear X-ray burst
observed from IGR J17591--2342 with {\it AstroSat}/LAXPC (see section~\ref{Model_reflection}).
The spectrum is for $8-10$ s since the burst onset and the XSPEC model
is {\tt tbabs(bbodyrad+constant*powerlaw)} with {\tt powerlaw} parameter
values frozen to the preburst values (section~\ref{color}).
{\it Upper panel}: Unfolded spectrum: data with $1\sigma$ errors are shown in red.
The total model (solid histogram), {\tt tbabs*bbodyrad} component (dashed line)
and {\tt tbabs*constant*powerlaw} component (dotted line) are also shown.
{\it Lower panel}:
Data to model ratio with errors.
This figure shows that this model is not adequate because there are two
excesses in $\sim 6-8$ keV and $\sim 12-20$ keV ranges.}
\label{figS1}
\end{figure}

\begin{figure}[ht]
\centering
\includegraphics*[width=8cm,angle=0]{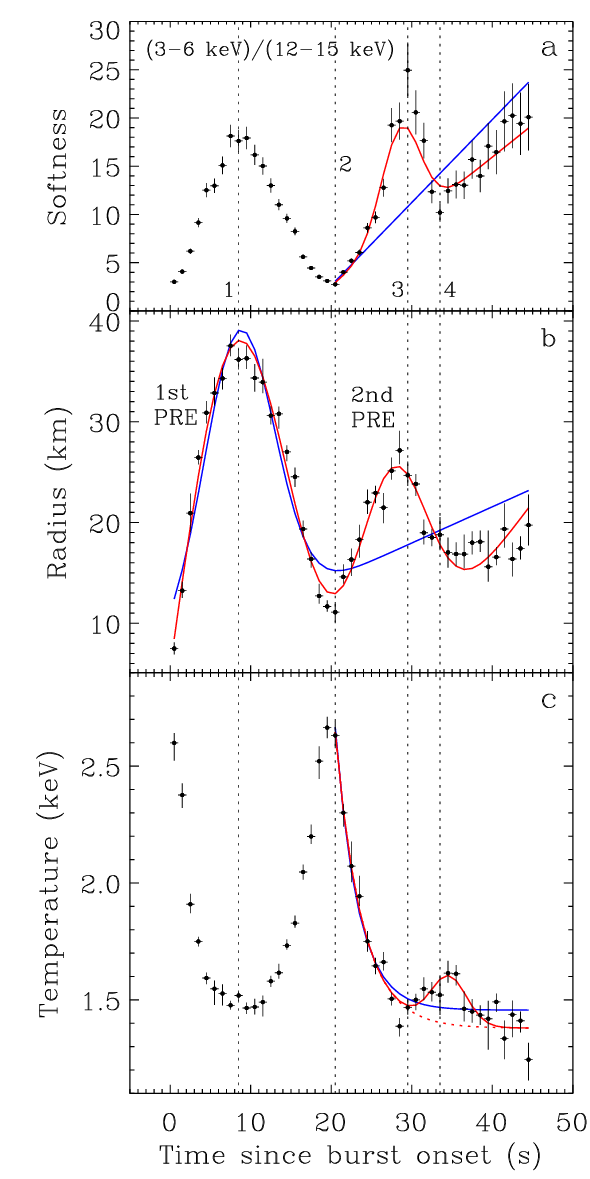}
        \caption{Time-resolved (1 s bins) properties, with $1\sigma$ errors,
        of the thermonuclear X-ray burst observed from IGR J17591--2342 with
        {\it AstroSat}/LAXPC (see section~\ref{significance}). 
	Parameter values of {\it panels a, b, c}
        are the same as {\it panels c, d, e}
        of Figure~\ref{fig2}, respectively, and vertical dotted lines mark the approximate
        maxima and minima of softness ({\it panel a}).
        The spectral parameter values of {\it panels b} and {\it c} are
        obtained by spectral fitting with the
        {\it XSPEC} model {\tt tbabs(bbodyrad+constant*powerlaw+gauss+gauss)}
        with {\tt powerlaw} parameter values frozen to the pre-burst values 
	(section~\ref{color}).
        In each panel, the blue best-fit model (solid line)
        does not include a {\it Gaussian} component for
        the hump likely due to a second PRE, while the red best-fit model (solid line) includes
        such a component. An F-test between these two models gives the
        significance of the hump likely due to a second PRE (see section~\ref{significance}).
        In {\it panel c}, the dotted red line is the solid red line minus its
        {\it Gaussian} component. Hence, a comparison between the solid and dotted
        red lines in {\it panel c} clearly shows the size of the
        blackbody temperature hump, likely due to a second PRE.}
\label{figS2}
\end{figure}

\subsection{Model with reflection and double-PRE interpretation}\label{Model_reflection}
 
Therefore, we examine the data-to-model ratio from spectral fitting to understand why $\chi^2_{\nu}$ values are high for the basic model, generally finding two excesses contributing the most in the $\sim 6-8$ keV and $\sim 12-20$ keV ranges of the
data-to-model ratio profile (see, e.g., Figure~\ref{figS1}). 
Fitting the two excesses with Gaussian components decreases the $\chi^2_{\nu}$(DoF) from 4.88(29) to 1.95(26) (Gaussian for $\sim 6-8$ keV), and finally to 1.22(23) (Gaussian also for $\sim 12-20$ keV).
Based on that, we update the model to {\tt tbabs(bbodyrad+constant*powerlaw+gauss+gauss)} for all burst
time bins to perform a uniform spectral analysis. However, we also verify our results with only the $\sim 6-8$ keV Gaussian
for those time bins for which the $\sim 12-20$ keV excess is not prominent. 

As shown in panel {\it g} of Figure~\ref{fig2}, the updated model provides $\chi^2_{\nu}$ values always lower than 2 (for 1 s burst time bins), $\sim 80\%$ of which are lower than 1.5. Considering a $\chi^2_{\nu}$ standard deviation of $\sim 0.3$ for a mean DoF of $\sim 17$, the limited statistics for the selected short intervals, and the modest spectral resolution of LAXPC, we can consider the fits acceptable and decide to include no further components.

Before presenting the evolution of blackbody parameter values,
let us discuss our interpretation of the other model components. As previously mentioned in Section~\ref{color}, the power law component represents the Comptonization component from the non-burst emission.
The flux of this component increases during the burst,
as can be seen from a significant increase of the {\tt constant} component (on average $\sim 27$ for burst time bins), likely reflecting an enhancement of the accretion rate caused by the radiation-induced
Poynting--Robertson drag on the accretion disc \citep{Walker1992, Worpeletal2013}. However, the direct disc emission
is not spectrally detected due to the limited LAXPC energy range at relatively low energies.

The two Gaussian components can be naturally described as features of the reflection spectrum off the accretion disc, 
with the $\sim 6-8$ keV component representing the Fe K$\alpha$ line, while the $\sim 12-20$ keV component is associated with the Compton back-scattering \citep{Ballantyne2004, Miller2007}. Such a reflection spectrum during thermonuclear bursts has been observationally reported earlier \citep{Guver2022}. We note that photons from both burst and Comptonization components should have been reflected off the disc; hence, a complex reflection model is required to describe it.  
We test that neither the model for burst photons' reflection off the accretion disc \citep[{\tt bbrefl in \it XSPEC}; ][]{Ballantyne2004} nor the model {\tt pexmon} ({\it XSPEC}) for reflection of photons from a Comptonization (cut-off power law) component gives acceptable fits for most of the burst time bins, indicating contributions from both may be required.

Now, let us consider the evolution of blackbody parameter values. Panels {\it d--e} of Figure~\ref{fig2} show the profiles of best-fit blackbody ({\tt bbodyrad}) parameters---radius ($r_{\rm BB}/D_{10}$) and temperature ($kT_{\rm BB}$).
The blackbody radius profile shows two consecutive humps which are aligned with temperature dips. This peculiar behaviour is consistent with two subsequent PRE events, each of which is compatible with the typical increase (decrease) and subsequent decrease (increase) of the blackbody radius (blackbody temperature) \citep[see, e.g.,][]{Gallowayetal2020}. However, while a PRE event during a strong thermonuclear burst is common, a second PRE during the same type-I X-ray burst is not typically expected or detected. Therefore, first, we need to establish the reliability of the second blackbody radius hump, aligned with a temperature dip.

\subsubsection{Reliability of the simultaneous second radius hump and temperature dip: model-independent evidence}\label{model-independent}

We start by comparing the evolution of the colours, which are spectral-model-independent, with respect to the blackbody parameters. The softness parameter (estimated in a spectral-model-independent way) reported in panel {\it c} of Figure~\ref{fig2} clearly resembles the blackbody radius evolution shown in panel {\it d} of the same figure. The behaviour also aligns with the blackbody temperature profile, resembling the expected PRE temporal evolution (panel {\it e} of Figure~\ref{fig2}). This consistency suggests that the obtained blackbody parameter profiles of IGR J17591--2342 are reliable. Moreover, we note that the bolometric burst intensity decreases near-exponentially since $\sim 20$ s after the burst onset
(panel {\it f} of Figure~\ref{fig2}; also Figure~\ref{fig1}). Hence, the blackbody temperature is expected to decrease overall. 
Therefore, assuming two consecutive PREs during the burst, the $kT_{\rm BB}$ maximum at the second touchdown point ($\sim 33 $ s after the burst onset) is not as high as that at the first touchdown point ($\sim 20 $ s after the burst onset). Nevertheless, a $kT_{\rm BB}$ increase on top of an overall decline is clearly seen, which is correlated with a similar
increase of the spectral-model-independent $12-15$ keV count rate and the decrease in softness between the vertical dotted lines 3 and 4 in Figure~\ref{fig2}. We interpret these as a supporting evidence that the second blackbody radius hump and the $kT_{\rm BB}$ minimum aligned with it, which could represent a second PRE event, does not originate from spectral modelling artefact.

\subsubsection{Reliability of the simultaneous second radius hump and temperature dip: significance}\label{significance}

We then proceed by estimating the significance of the second blackbody radius hump and simultaneous temperature and colour features.
We fit the entire blackbody radius profile
(panel {\it b} of Figure~\ref{figS2}) with two models:
a {\it linear+Gaussian} and a {\it linear+Gaussian+Gaussian}, where the first
{\it Gaussian} is for the first radius hump and the second {\it Gaussian} is for
the second radius hump. Fitting with the first model gives $\chi^2$(DoF) $= 602.0(40)$,
and that with the second model gives $\chi^2$(DoF) $= 88.3(37)$.
Thus an F-test between these two models shows that the second radius hump is
$\approx 8\sigma$ significant.
If there were no second PRE,
a near-exponential decay of the blackbody temperature ($kT_{\rm BB}$) after
the (first) touchdown point could be expected
(but, see section~\ref{PRE_atmosphere}). But, if the second PRE occurred,
then $kT_{\rm BB}$ should have increased during the blackbody radius decrease,
thus having a maximum at the second touchdown point. We indeed find such a
hump, and in order to find its significance, we fit the
$kT_{\rm BB}$ profile from the
first touchdown point with an {\it exponential+constant} model and
an {\it exponential+constant+Gaussian} model
(panel {\it c} of Figure~\ref{figS2}),
where the {\it Gaussian} is for the $kT_{\rm BB}$ hump. Fitting with the first model gives
$\chi^2$(DoF) $= 69.3(22)$, and that with the second model gives $\chi^2$(DoF)
$= 31.5(19)$. Therefore, an F-test between these two models shows that the
$kT_{\rm BB}$ hump is $\approx 3.2\sigma$ significant.
We also fit the softness profile from the first touchdown point with a
{\it linear} model and a {\it linear+Gaussian} model
(panel {\it a} of Figure~\ref{figS2}),
and find with an F-test that the second softness hump, described by the {\it Gaussian},
is $\approx 5.2\sigma$ significant.

\subsubsection{Reliability of the simultaneous second radius hump and temperature dip: robustness}\label{robustness}

Finally, we also suggest the robustness of (i) the second blackbody radius hump aligned with a $kT_{\rm BB}$ dip, and (ii) a subsequent $kT_{\rm BB}$ peak aligned with a low radius value, by noticing that they do not depend on any of the following factors: (i) the evolution of the non-burst spectral shape during the burst, (ii) a specific pre-burst spectral model, (iii) whether the burst spectra are of 1 s or 2 s duration, (iv) and fitting of the excesses in $\sim 6-8$ keV and $\sim 12-20$ keV ranges.

In order to show these, we create burst spectra for both 1 s and 2 s
time bins and fit them with various reasonable models.
First, we check if the inclusion of a non-burst spectral shape evolution during
the burst provides a better fit and affects the second radius hump signature.
The upper panel of Figure~\ref{figS0} clearly shows the burst data do not require a non-burst spectral shape evolution for the basic spectral model having only absorbed blackbody and power law.
However, the fits for most spectra are unacceptable for this model
due to high $\chi^2_{\nu}$ values.

\begin{figure*}
\centering
\hspace{0.0cm}
\includegraphics*[width=14.0cm,angle=0]{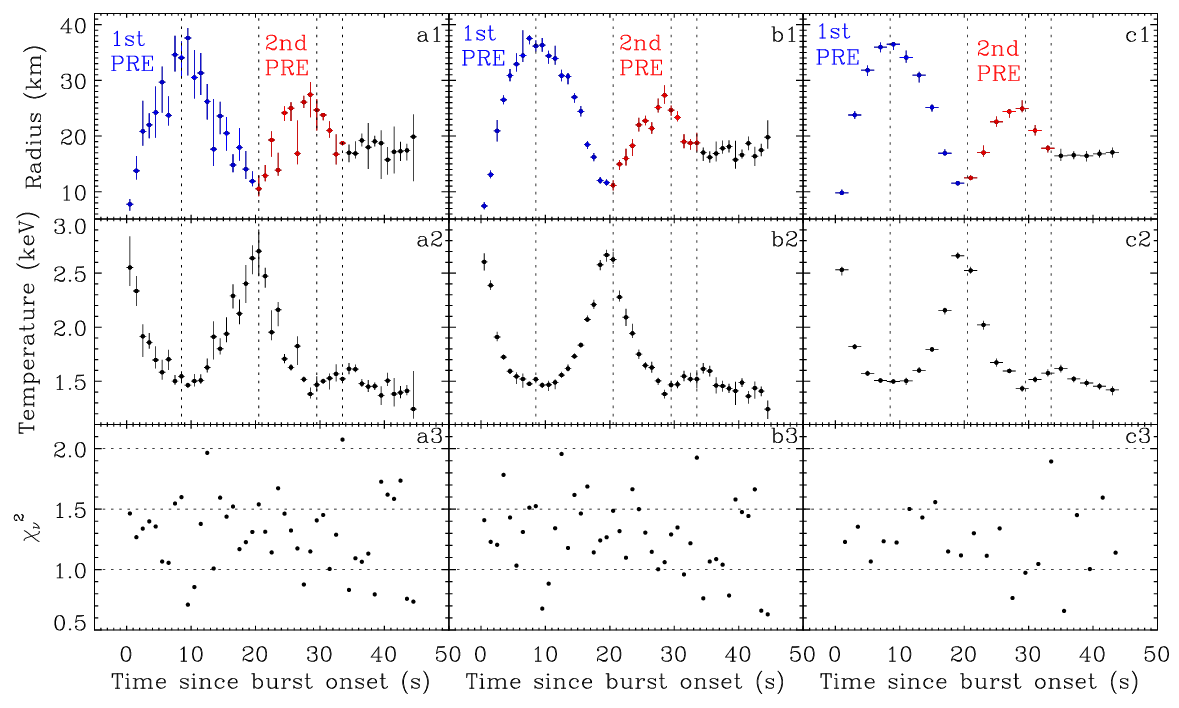}
       \caption{Time-resolved properties, with $1\sigma$ errors,
        of the thermonuclear X-ray burst observed from IGR J17591--2342 with
        {\it AstroSat}/LAXPC (see section~\ref{robustness}).
Blackbody parameter (radius and temperature) and reduced $\chi^2$ profiles are shown for three cases:
(i) fitting of 1 s burst spectra with the {\it XSPEC} model {\tt tbabs(bbodyrad+powerlaw+gauss+gauss)} ({\it panels a1, a2, a3});
(ii) fitting of 1 s burst spectra with the {\it XSPEC} model {\tt tbabs(bbodyrad+constant*Nthcomp+gauss+gauss)} ({\it panels b1, b2, b3}); and
(iii) fitting of 2 s burst spectra with the {\it XSPEC} model {\tt tbabs(bbodyrad+constant*powerlaw+gauss+gauss)} (the second Gaussian in $\sim 12-20$~keV is included if required; {\it panels c1, c2, c3}). For (ii) and (iii),
Comptonization ({\tt Nthcomp}) and power law parameter values are
frozen to the preburst values (section~\ref{color}).
This figure shows that the second blackbody radius hump aligned with a temperature dip is present
for multiple reasonable models and spectra for different time bin sizes,
and hence is robust.}
\label{figS3}
\end{figure*}

From Figure~\ref{figS1}, we find that the high $\chi^2_{\nu}$ values are due
to the $\sim 6-8$~keV and $\sim 12-20$~keV excesses.
Therefore, we include two Gaussians to describe these excesses in
our updated spectral model (see section~\ref{Model_reflection}).
Thus, in order to test if the inclusion of a non-burst spectral shape
evolution during the burst affects the second radius hump signature,
we include these two Gaussians and allow the non-burst spectral shape to
evolve (see panels {\it a1}, {\it a2}, and {\it a3} of
Figure~\ref{figS3}).
Comparing this figure with Figure~\ref{fig2}, for which the spectral model
is the same except the non-burst spectral shape is frozen at the pre-burst value,
we find that the $\chi^2_{\nu}$ values are generally not improved for the
former. This suggests that a non-burst spectral shape evolution
is not required by the burst data, even when the excesses are fit with a suitable
models, and the fits are acceptable.

From panels {\it a1} and {\it a2} of Figure~\ref{figS3}, we find that the burst blackbody radius and temperature profiles
are very similar to those for frozen non-burst spectral shapes (Figure~\ref{fig2}), although, for the former, as expected, the error bars are larger.
As before (see section~\ref{significance}), We fit the blackbody radius profile with a {\it linear+Gaussian}
($\chi^2$(DoF) $= 237.1(40)$) and a {\it linear+Gaussian+Gaussian}
($\chi^2$(DoF) $= 43.3(37)$), which implies from an F-test that
the second radius hump is $\approx 7.4\sigma$ significant.
This shows that the second radius hump is not an artefact due to the
freezing of the non-burst spectral shape during the burst.

The pre-burst spectrum can also be fit with a physical Comptonization
model ({\it XSPEC} model {\tt Nthcomp}), as mentioned in section~\ref{color}.
Therefore, we fit the burst spectra with the {\it XSPEC} model \\
{\tt tbabs(bbodyrad+constant*Nthcomp+gauss+gauss)},
where {\tt Nthcomp} parameter values are frozen to the preburst values.
The burst blackbody radius and temperature profiles clearly show the
signature of the second radius hump aligned with temperature dip,
and the $\chi^2_{\nu}$ values are similar to those for the {\it XSPEC} model \\
{\tt tbabs(bbodyrad+constant*powerlaw+gauss+gauss)}
(see panels {\it b1}, {\it b2}, and {\it b3} of
Figure~\ref{figS3}, and Figure~\ref{fig2}).
This suggests that the second radius hump and the aligned temperature dip
do not depend on the specific model used to fit the preburst spectrum.

In order to check if the specific nature of the blackbody radius and temperature
profiles depend on the 1 s duration of burst spectra, we make spectra for 2 s
independent time bins during the burst and fit them with the {\it XSPEC}
model {\tt tbabs(bbodyrad+constant*powerlaw+gauss+gauss)}, with
power law parameter values frozen to the pre-burst values
(the second Gaussian in $\sim 12-20$ keV is included if required;
see panels {\it c1}, {\it c2} and {\it c3} of Figure~\ref{figS3}).
The $\chi^2_{\nu}$ values are similar to those for 1 s burst spectra
(compare with Figure~\ref{fig2}).
The blackbody radius and temperature profiles for 2-second spectra clearly
show the second radius hump and the aligned temperature dip, which do not depend on whether we use
both Gaussians in the spectral model, or, as required,
only the one in $\sim 6-8$ keV.

If we compare panels {\it c1} and {\it c2} of Figure~\ref{figS3}
with the middle and lower panels of Figure~\ref{figS0},
respectively, we find similar blackbody radius and temperature profiles,
and hence a clear signature of the second radius hump in both, although
Gaussian models are not included for the latter figure.
Therefore, the signature of the second radius hump, aligned with the temperature dip, is independent of the Gaussian
models used to describe the $\sim 6-8$~keV and $\sim 12-20$~keV excesses.

Finally, note that, while the fast spin and the resulting oblateness of the neutron star could affect the obtained spectral parameter values by
a few percent \citep{Suleimanovetal2020}, they cannot give rise to the
humps and dips in the observed blackbody parameter profiles.

\subsection{Model with spectral edge and an interpretation of evolving stellar atmospheric composition}\label{PRE_atmosphere}

Here, we fit the burst spectra with an alternative model for reasons 
explained below, and compare its suitability relative to the 
spectral model described in subsection~\ref{Model_reflection}.

We show that the burst blackbody radius ($r_{\rm BB}/D_{10}$) profile
for IGR J17591--2342 has a
significant and robust second hump simultaneously with a blackbody temperature
($kT_{\rm BB}$) dip (see subsections~\ref{Basic_model}, \ref{Model_reflection}).
From this, one could conclude the occurrence of a second PRE during the burst,
assuming that the photospheric radius $r$ and the
effective temperature $kT_{\rm eff}$ behave like $r_{\rm BB}/D_{10}$ and
$kT_{\rm BB}$, respectively.
But, there is a way that $r_{\rm BB}/D_{10}$ and $kT_{\rm BB}$ could show such an
anti-correlated behaviour, even when the photospheric radius $r$
does not change, e.g.,
the photosphere remains on the neutron star surface ($r=R$; $R$: stellar radius).
This can be seen from section~\ref{Basic_model}.
If $r$ does not change, $z_r$ remains the same.
$D_{10}$ is, anyway, a constant for the source. Therefore, in this case,
$r_{\rm BB}/D_{10} \propto f_{\rm c}^{-2}$. Thus, if $f_{\rm c}$ temporarily
decreases during a burst, the blackbody radius can temporarily increase
even if the photospheric radius $r$ remains the same.
A temporary decrease of $f_{\rm c}$ can also temporarily reduce
$kT_{\rm BB}$.
On the other hand, if $f_{\rm c}$ increases, $R_{\rm BB}/D_{10}$ can
decrease and $kT_{\rm BB}$ can increase.
Therefore, the signature of the second PRE, and even that of the first PRE,
could result from a systematic evolution of $f_{\rm c}$.

What can cause a significant decrease and subsequent increase of
$f_{\rm c}$ during one or more periods of a thermonuclear burst?
This is possible if the metallicity of the neutron star atmosphere
significantly increases and then
decreases during such a period \citep{Suleimanovetal2011,Kajavaetal2017}.
To model such phenomena, ideally, one needs to use realistic spectra of stellar atmospheres with various chemical compositions for a
static/expanding/contracting photosphere.
There have been many efforts to compute realistic atmospheric
spectra \citep[e.g., ][]{Londonetal1986, Madej1991, Suleimanovetal2011}.
However, a general atmospheric model with various compositions, considering photospheric expansion, is not yet publicly available for X-ray spectral fitting (e.g., using {\it XSPEC}). 
On the other hand, a diluted blackbody model
with a parameter $f_{\rm c}$, used to correct the measured photospheric radius
and temperature, usually fits the burst spectra well and is popularly used
\citep[e.g., ][]{Paradijs1990,Suleimanovetal2011,Keeketal2018,Suleimanovetal2020}.
We use this model in this paper.
However, only from this blackbody fitting it is not possible to discriminate
between an atmosphere with evolving metal content (which should cause an
$f_{\rm c}$ evolution) and an expanding/contracting photosphere.
But another way to infer a metal content evolution is available.
The spectrum from a metal-rich stellar atmosphere is typically expected
to contain an absorption edge (around $7-10$ keV) due to partially ionised
Fe/Ni \citep[e.g., ][]{Weinberg2006,ZandWeinberg2010,Kajavaetal2017,Lietal2018}.
Thus, one can test if the models used to fit observed burst spectra
require such an edge.

\begin{figure}[ht]
\centering
\includegraphics*[width=7.0cm,angle=0]{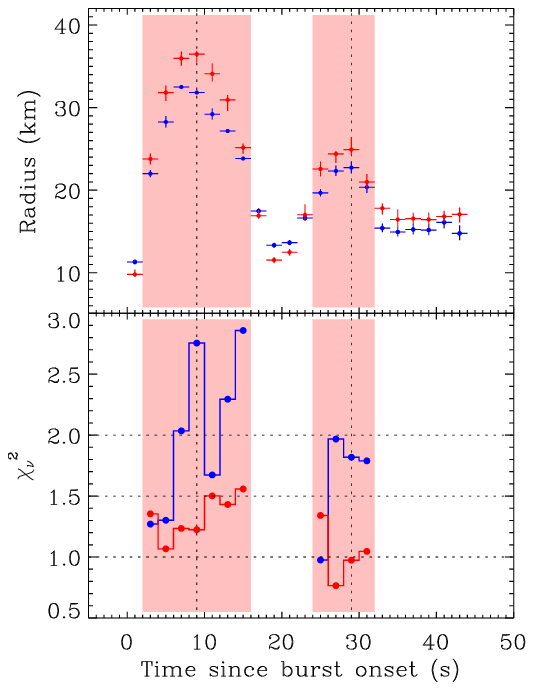}
       \caption{Time-resolved properties, with $1\sigma$ errors,
        of the thermonuclear X-ray burst observed from IGR J17591--2342 with
        {\it AstroSat}/LAXPC (see section~\ref{PRE_atmosphere}).
The upper panel shows the blackbody radius profiles for spectral fitting with the {\it XSPEC} models {\tt tbabs(bbodyrad+constant*powerlaw+gauss+gauss)} (red; the second Gaussian in $\sim 12-20$~keV is included if required)
and {\tt tbabs(edge*bbodyrad+constant*powerlaw)} (blue).
The lower panel shows the reduced $\chi^2$ values for the same spectral models
(marked in same colours) for two phases of radius increase
(highlighted in pink
patches). The vertical dotted lines mark the approximate radius maxima.
This figure shows that during blackbody radius humps the model including Gaussian(s) overall fits the burst spectra much better than the model including edge, but the radius profiles are qualitatively the same for
both models.
}
\label{figS4}
\end{figure}

\begin{figure}[ht]
\centering
\hspace{0.0cm}
\includegraphics*[width=8.5cm,angle=0]{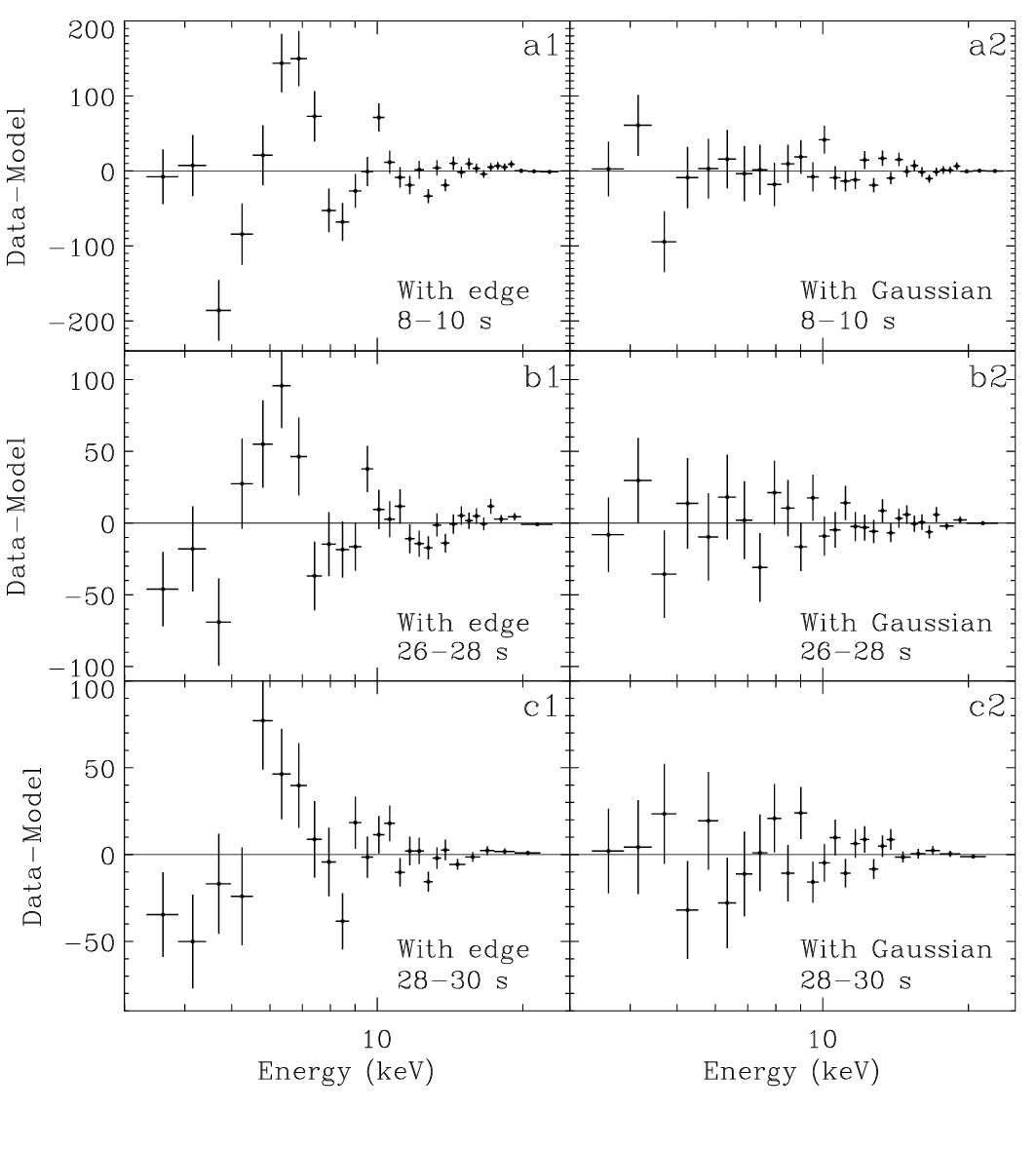}
       \caption{Residuals (data$-$model) of fitting of the thermonuclear X-ray burst spectra (of 2 s time durations) observed from IGR J17591--2342 with {\it AstroSat}/LAXPC (see section~\ref{PRE_atmosphere}).
Three spectra for $8-10$ s ({\it panels a1, a2}), $26-28$ s ({\it panels
b1, b2}) and $28-30$ s ({\it panels c1, c2}) since the burst onset are
considered. These are times during the blackbody radius peaks.
Left panels ({\it a1, b1, c1}) are for the {\it XSPEC} spectral model
{\tt tbabs(edge*bbodyrad+constant*powerlaw)}, and right panels ({\it
a2, b2, c2}) are for the model
{\tt tbabs(bbodyrad+constant*powerlaw+gauss+gauss)} (the second Gaussian in $\sim 12-20$~keV is included if required).
This figure shows that during radius peaks, the model including
Gaussian(s) fits the burst spectra better than the model including edge does.
}
\label{figS5}
\end{figure}

Therefore, let us now look into our spectral results and perform additional spectral fits, including an absorption edge component, to critically examine the alleged PREs.
As previously mentioned, Figure~\ref{figS0} shows two
periods of unacceptable fits during the burst for the basic model components,
blackbody (for burst emission) and power law (for non-burst emission).
These periods coincide with the blackbody radius humps, i.e., the alleged PREs.
For both these periods, unacceptable fits are due to unmodelled excesses
in $\sim 6-8$~keV and $\sim 12-20$~keV (see Figure~\ref{figS1};
only in $\sim 6-8$~keV for some burst time bins). This similarity in spectra
could indicate that the blackbody radius humps for both periods occur for
the same reason---either an $f_{\rm c}$ evolution due to a metallicity
evolution, or a photospheric expansion/contraction.
After the excesses are fit with Gaussians, a spectral absorption edge is
not required, indicating no significant metallicity of the stellar atmosphere.
However, could these excesses appear due to the lack of an absorption
edge model component and not the absence of Gaussian components?
Indeed, such a spectral edge component (without a Gaussian) could
sufficiently improve the fit for bursts from other
sources \citep{Kajavaetal2017,Lietal2018}.

To test this, we fit the burst spectra with
{\tt tbabs(edge*bbodyrad+constant*powerlaw)}.
In this {\it XSPEC} model, the multiplicative component {\tt edge}
(absorption edge) as a function of energy ($E$) is
$1$ for $E < E_{\rm c}$ and
$\exp[-D(E/E_{\rm c})^{-3}]$ for $E \ge E_{\rm c}$.
Here, $E_{\rm c}$ is the threshold energy and
$D$ is the absorption depth at the
threshold\footnote{https://heasarc.gsfc.nasa.gov/xanadu/xspec/manual/node242.html}.
The {\tt edge} should originate from the neutron star atmosphere and hence is
multiplied to {\tt bbodyrad} (blackbody).
Figure~\ref{figS4} compares the blackbody radius profiles
and $\chi^2_{\nu}$ values during the radius humps between two cases:
when the $\sim 6-8$~keV and $\sim 12-20$~keV excesses are modelled with
an absorption edge, and when they are modelled with Gaussian(s).
First, we find that the radius humps are qualitatively independent of the model component used to describe the excesses in the spectrum.
This further shows the robustness of the second radius hump
(see section~\ref{robustness}).
This figure also shows that during blackbody radius humps the model including Gaussian(s) overall fits the burst spectra much better
than the model including an absorption edge.
For example, for the 2 s spectrum at the first blackbody radius peak
($8-10$ s since the burst onset), $\chi^2$(DoF) $\approx 74.4(27)$ for the model including edge. The probability ($P$) for a value drawn
from the $\chi^2$ distribution to be equal to or higher than this obtained
$\chi^2$ value for the given DoF is $\approx 2.6\times10^{-6}$.
But, for the same spectrum and the model including Gaussian(s),
$\chi^2$(DoF) $\approx 28.2(23)$ and $P \approx 0.21$.
This shows that the model including Gaussian(s) fits significantly better
than the model including an edge.

For the two 2-s spectra at the second blackbody radius peak
($26-28$ s and $28-30$ s since the burst onset), $\chi^2$(DoF) $\approx 47.2(24)$
and $\approx 36.4(20)$, respectively, for the model including edge.
However, for these spectra, $\chi^2$(DoF) $\approx 15.3(20)$
and $18.5(19)$, respectively, for the model including Gaussian(s).
Since all the 2 s spectra are for independent time bins, we add the
$\chi^2$ values and DoF values separately for the four spectra during
the second blackbody radius hump (the second pink patch in
Figure~\ref{figS4}).
For these four spectra, $\chi^2$(DoF) $\approx 148.5(93)$ and $P \approx 0.0002$
for the model including an edge, and
$\chi^2$(DoF) $\approx 86.9(83)$ and $P \approx 0.36$
for the model including Gaussian(s).
Therefore, although the observed flux during the second radius hump is
much smaller than that during the first radius hump, the two models
can be discriminated during the second hump and the model including Gaussian(s)
fits significantly better.
Note that if we combine the $\chi^2$ values and DoF values separately for all the 22 burst spectra of 2 s, we find
$\chi^2$(DoF) $\approx 894.5(559)$ and $P \approx 6.4\times10^{-18}$
for the model including an edge, and
$\chi^2$(DoF) $\approx 613.1(495)$ and $P \approx 0.0002$
for the model including Gaussian(s).
This suggests that the latter model overall fits the burst spectra
significantly better than the former model.

To gain further insight, we also present the residuals of fits for the two models for the three above-mentioned burst spectra, one at
the first blackbody radius peak and two at the second blackbody
radius peak (see Figure~\ref{figS5}). This figure
clearly shows that a Gaussian can sufficiently fit the $\sim 6-8$~keV
excess, while an absorption edge cannot.
Thus, an absorption edge cannot explain the burst spectra either for the first blackbody radius peak or for the second radius peak.
This implies that an atmospheric metallicity, and hence $f_{\rm c}$,
evolution may not cause the blackbody radius humps and therefore these humps,
and the aligned blackbody temperature dips,
could be due to neutron star photospheric expansions.

\section{Discussion and interpretation}\label{Discussion}

In this paper, we study an unusual type-I X-ray burst from the neutron
star X-ray binary and accretion-powered millisecond X-ray pulsar 
IGR J17591--2342. A detailed time-resolved spectroscopic investigation
of the burst shows two significant and robust subsequent blackbody
radius humps aligned with two blackbody temperature dips. This could be
interpreted as two PREs during the same burst, or, the second 
radius hump could be due to the neutron star atmospheric 
metallicity evolution. Each of these scenarios is extremely interesting.
While the first one implies that a significant amount of fuel in the 
vicinity of the burning region of a powerful burst can remain unburnt,
the second one could enable neutron star gravitational redshift
measurement through observations of spectral lines.
Let us first compare the reliabilities of these two scenarios.

\subsection{A discussion on two scenarios}\label{Discussion_two_scenarios}

The second scenario, i.e., the neutron star atmospheric metallicity evolution, implies that a metal-rich stellar atmosphere should show 
an absorption edge component in the spectrum (around $7-10$ keV) due to partially ionized Fe/Ni \citep[e.g., ][]{Lietal2018}.
But the fitting of burst spectra with the second model involving a spectral edge is clearly much worse compared to that with the first model involving reflection and interpreted as the double-PRE
(see section~\ref{PRE_atmosphere}).
Nevertheless, a metal-rich atmosphere without a strong edge is still plausible, for example, due to a relatively low Fe/Ni abundance. Therefore, while the occurrence of the first PRE is not unexpected, one cannot entirely rule out an atmospheric metallicity evolution origin of the second PRE signature.

Nonetheless, our results favour
a genuine PRE origin of the second blackbody radius 
hump, aligned with a temperature dip, for the following reasons. 
(i) Several previous papers predicted or found a spectral edge suggesting a metal-rich atmosphere \citep{Weinberg2006, ZandWeinberg2010, Kajavaetal2017, Lietal2018}, which we do not find. 
(ii) Spectral excesses in $\sim 6-8$ keV and $\sim 12-20$ keV ranges during both the burst blackbody radius humps indicate similar physical processes. Hence, a PRE origin of the first hump may suggest the same origin of the second hump. 
Otherwise, it is difficult to explain why there was a significant reflection 
off the disc during the first radius hump when the photosphere expanded,
and during the second radius hump when the photosphere did not expand,
but no significant disc reflection in between two radius humps when the photosphere did not expand.
(iii) The lack of a spectral absorption edge during the first radius hump could suggest a metal-poor atmosphere, with minimal or null enrichment even during a PRE event. Hence, explaining a subsequent influx of metals sufficient to produce a second blackbody radius hump bacomes difficult to explain coherently. 

\subsection{Burst flux evolution and its interpretation }\label{flux}

Here, we present the flux evolution of the burst observed from IGR J17591--2342, and, considering the above, interpret this evolution assuming two PRE events during the burst (see section~\ref{burst}). For each 1 s burst time bin, we estimate the bolometric ($0.01-100$ keV) flux ($F_{\rm BB}$) of the burst blackbody ({\tt bbodyrad}) component using the {\it XSPEC} {\tt cflux} model (see panel {\it f} of Figure~\ref{fig2}).
$F_{\rm BB}$ relates to the luminosity at the observer by $L_{\rm BB} = 4 \pi D^2 \xi F_{\rm BB}$,  where $D$ is the source distance and $\xi$ is the anisotropy factor due to mass distribution (e.g., accretion disc, Comptonizing hot electron gas) around the neutron star \citep[e.g., ][]{Kuiperetal2020}. On the other hand, the Eddington luminosity at the observer for a spherically symmetric burst emission is
$L_{\rm BB, Edd} = [4 \pi G M c]/[0.2 (1+X) (1+z_R)]$  \citep[e.g., ][]{ShapiroTeukolsky2004, Bultetal2019}, where $c$ is the speed of light in vacuum and $X$ is the fractional hydrogen abundance by mass in atmospheric layers.
Note that a significantly asymmetric burning on the spinning neutron star would likely generate burst oscillations \citep{Watts2012}, the lack of which suggests a spherically symmetric burst emission (see appendices~\ref{burst_oscillations} and \ref{burst_rise}). 
As shown in Figure~\ref{fig2}, $F_{\rm BB}$ first increased sharply within a second (consistent with Figure~\ref{fig1}). Then it remained at a similar level for the first $\sim 20$ s, when the first PRE occurred, although there is an indication of a dip at $\sim 10$ s since the burst onset. We measure a maximum $F_{\rm BB}$ of $\approx 7.5\times10^{-8}$ erg s$^{-1}$ cm$^{-2}$ \citep[note that][reported a burst peak $F_{\rm BB}$ of $\sim 7.6\times10^{-8}$ 
erg s$^{-1}$ cm$^{-2}$ for their burst from this source]{Kuiperetal2020}. Finally, $F_{\rm BB}$ decreased near-exponentially when the second PRE occurred.

$L_{\rm BB, Edd}$, should correspond to the touchdown $F_{\rm BB}$ \citep{Damenetal1990, Ozel2006}, however, there are two widely different touchdown $F_{\rm BB}$ values, $\approx 6.3\times10^{-8}$ erg s$^{-1}$ cm$^{-2}$ for the first and $\approx 2.0\times10^{-8}$ erg s$^{-1}$ cm$^{-2}$ for the second PRE events, implying a mismatch of a factor of $\approx 3$.
Equating $L_{\rm BB, Edd}$ to $L_{\rm BB}$, we find that $F_{\rm BB}$ can be different between the two touchdown points if $X$ and/or $\xi$ are different, because $F_{\rm BB} \propto (1+X)^{-1} \xi^{-1}$. $X$ can range from 0 (pure helium) to 0.73 (cosmic abundance of hydrogen), which means that $1+X$ could, in principle, explain up to a mismatch of a factor of 1.73.
A substantial increase in the anisotropy factor $\xi$ between the two touchdown points should be accompanied by an irregular variation in the observed flux \citep{MeliaZylstra1992}. However, we observe a relatively smooth, near-exponential decay (Figures~\ref{fig1} and \ref{fig2}), similar to most other thermonuclear bursts. Note that it is unlikely that three largely independent parameters, $kT_{\rm BB}$, $X$, and $\xi$, evolved in a coordinated way to give rise to such a regular burst decay profile. Moreover, $\xi$ could not have continued to increase monotonically from a less than 1 value \citep[e.g., $\xi = 1$ implies isotropy; ][]{Kuiperetal2020}. After its maximum value, $\xi$ would have either decreased or
remained the same, and hence, the burst decay profile should have become significantly less steep abruptly.
Thus, $X$ and $\xi$ may not explain two widely different
touchdown $F_{\rm BB}$ values.

\subsection{A discussion on the double-PRE scenario}\label{Discussion_double-PRE}

Considering the double-PRE scenario, our analysis suggests the possibility of two novel and unexplained aspects of the thermonuclear
type-I X-ray burst from IGR J17591--2342: (i) two PRE events within a few seconds and (ii) two touchdown luminosities  (likely Eddington luminosities) different by a factor of $\approx 3$ (see sections~\ref{burst} and \ref{flux}).
Conscious of the theoretical challenge to explain two consecutive PRE events during the same type-I X-ray burst due to the need to 
preserve a significant amount of fuel, sufficient to cause a second PRE event a few seconds after a strong burning with the first PRE, we will briefly present a speculative, albeit reasonable, and skeletal explanation for the above-mentioned puzzling findings.

The sharp burst rise and a subsequent PRE during a plateau-type $F_{\rm BB}$ profile (see section~\ref{flux})
strongly suggests an almost pure helium ignition and burning \citep[e.g., ][]{StrohmayerBildsten2006,Gallowayetal2020,Bultetal2019,Jaisawaletal2019}.
However, the $F_{\rm BB}$ plateau phase for a helium burst is usually much shorter \citep{StrohmayerBildsten2006} than the observed $\sim 20$ s duration of this phase of the current burst, and indeed a previous thermonuclear burst observed from this source showed a plateau duration of $\sim 10$ s (see section~\ref{introduction}). Combining this aspect with the previously mentioned dip in the flux profile $\sim 10$ s after the burst onset, and the long duration of the burst suggests a second step of energy generation, perhaps by a subsequent thermonuclear burning of mixed hydrogen and helium stored at an upper layer \citep{Fujimotoetal1988}.
Hydrogen in the upper layer is compatible with the evidence of the companion star being a hydrogen-rich low-mass star \citep{Kuiperetal2020, Gusinskaiaetal2020}. However, it is unclear how this layer can remain unburnt during intense burning with the first PRE \citep{Fujimotoetal1988, Spitkovskyetal2002}, which may stimulate further detailed studies.

If the flux ($F_{\rm BB,td2}$) of the second touchdown point corresponds to $L_{\rm BB, Edd}$, then both the PREs, during which the
flux was higher (see Figure~\ref{fig2}), could naturally occur. Then, the remaining puzzle is why the photosphere touched down on the stellar surface after the first PRE ($\sim 20$ s after the burst onset) when
$F_{\rm BB}$ was $\approx 3$ times $F_{\rm BB,td2}$. As shown earlier, evolutions of $X$ and $\xi$ cannot explain this puzzle,
at least not entirely (section~\ref{flux}). An explanation could come from the magnetic confinement of atmospheric layers if the stellar magnetic field is tangled and enhanced by convection at high luminosities \citep{Boutloukosetal2010}, compatible with the fact that IGR J17591--2342, being an AMXP, has a relatively higher magnetic field \citep{Sannaetal2020, Tseetal2020}.
This convection could also mix helium in the atmosphere, thus decreasing $X$ and increasing $L_{\rm BB, Edd}$, and therefore helping the first touchdown. Note that, as the burst luminosity decreased $\sim 20$ s after the burst onset, perhaps helium largely subsided and magnetic field tangling reduced, favouring the second PRE event.

Finally, the likely preservation of a substantial fuel mass in the vicinity of intense nuclear burning and a strong suggestion of magnetic confinement of plasma during thermonuclear fusion can challenge the current theoretical understanding \citep[see, e.g.,][]{Fujimotoetal1988, Spitkovskyetal2002} and could be interesting for multiple fields such as plasma physics including dynamo mechanism, supernova physics, and the nuclear, and particle physics of neutron stars, which involves strong gravity, high density, large magnetic field and intense radiation.

\begin{acknowledgements}
The authors acknowledge {\it AstroSat} Large Area X-ray Proportional Counter (LAXPC) teams (TIFR), Indian Space Research Organisation (including the Indian Space Science Data Centre), and the High Energy Astrophysics Science Archive Research Center (NASA/GSFC). The authors also thank H. M. Antia and T. Katoch (LAXPC team) for discussion on LAXPC pointing.
\end{acknowledgements}

\bibliographystyle{aa}
\bibliography{ms}

\begin{appendix}

\section{A search for burst oscillations}\label{burst_oscillations}

We search for burst oscillations (see section~\ref{introduction})
in the frequency range of $524-530$ Hz (since
the neutron star spin frequency is $527.43$ Hz for IGR J17591--2342) and 
the energy range of $\approx 3-25$ keV throughout the first 45 s of the 
type-I burst. We apply a sliding window search
with window sizes of 1, 2, and 4 s \citep{Bultetal2019} with a step of 0.1 s.
Using the fractional amplitude maximisation technique \citep{ChakrabortyBhattacharyya2014}, which was shown to be similar to the $Z^2$ power maximisation technique \citep{StrohmayerMarkwardt2002}, we look for
at least $3\sigma$ detection of burst oscillations in each time bin. Although
a few time bins have better than $3\sigma$ detection, not even more than four
consecutive and overlapping bins have ever had such a detection. Therefore, we consider these
detections spurious. 
However, we estimate the upper limits of fractional
root-mean-squared (rms) amplitude, which, for independent time bins of 1 s,
comes out to be typically $\sim 3-4$\% for higher burst intensities and goes up to $\sim 13$\% in the burst tail.

\section{A study of the burst rise}\label{burst_rise}

The $3-50$ keV count rate during the burst from IGR J17591--2342 
reached close to 6000 in the
first $\approx 0.4$ s, and exceeded 10000 in the first $\sim 1.3$ s, which
implies a drastic slope change of the intensity profile
at $\approx 0.4$ s since the burst onset.
We perform a spectral fitting for the time period of
$\approx 0.25-0.55$ s since the burst onset,
and find a blackbody bolometric ($0.01-100$ keV) flux $F_{\rm BB}$ of
$\sim 2.6\times10^{-8}$ erg s$^{-1}$ cm$^{-2}$ using the
{\it XSPEC} {\tt cflux} model.
This is more than the $F_{\rm BB}$ value at the second touchdown point, which is
$\approx 2.0\times10^{-8}$ erg s$^{-1}$ cm$^{-2}$ (see panel {\it f} of Figure~\ref{fig2}).
Thus, the first PRE could have started
within the first $\sim 0.4$ s since the burst onset.
Considering the first 1/3 s of the burst rise, we find the
blackbody radius, temperature, bolometric flux, and the upper limit of the
burst oscillation fractional rms amplitude to be $3.4_{-1.9}^{+2.0}$ km,
$2.05_{-0.39}^{+0.50}$ keV, $\sim 2.2\times10^{-9}$ erg s$^{-1}$ cm$^{-2}$
and $\sim 33$\%, respectively.
Since this temperature is already consistent with the peak temperature
(see panel {\it e} of Figure~\ref{fig2}), the burning timescale could
be less than $\sim 1/3$ s, supporting a pure helium or significantly
hydrogen deficient burning \citep{StrohmayerBildsten2006} inferred in this paper.
The low value of the blackbody radius
suggests a localised ignition and that the thermonuclear flame had not yet spread
all over the neutron star surface (section~\ref{introduction}).
A relatively high upper limit of the burst oscillation fractional rms amplitude during the first $\sim 1/3$ s is consistent with this low value of the blackbody
radius (and hence a small burning region).
The low bolometric flux implies that the above-mentioned
Eddington flux was not reached within the first $\sim 1/3$ s.
Thus, flame spreading and PRE perhaps happened
simultaneously during the later part of the burst rise.

\end{appendix}

\end{document}